\documentclass{article}%
\usepackage{amssymb}
\usepackage{amsfonts}
\usepackage{amsmath}
\usepackage{graphicx}%
\begin{document}

\title{A Novel Solution to the ATT48 Benchmark Problem}
\author{Anthony A. Ruffa\\Naval Undersea Warfare Center Division\\1176 Howell Street\\Newport, RI 02841}
\maketitle

\begin{abstract}
A solution to the benchmark ATT48 Traveling Salesman Problem (from the
TSPLIB95 library) results from isolating the set of vertices into ten
open-ended zones with nine lengthwise boundaries. \ In each zone, a
minimum-length Hamiltonian Path (HP) is found for each combination of boundary
vertices, leading to an approximation for the minimum-length Hamiltonian Cycle
(HC). \ Determination of the optimal HPs for subsequent zones has the effect
of automatically filtering out non-optimal HPs from earlier zones. \ Although
the optimal HC for ATT48 involves only two crossing edges between all zones
(with one exception), \ adding inter-zone edges can accomodate more complex problems.

\end{abstract}

\section{Introduction}

Given a set of vertices, the well-known Traveling Salesman Problem (TSP)
involves finding the minimum-length Hamiltonian Cycle (HC): the path visiting
each vertex once and returning to the starting vertex.

The symmetric TSP with $N$ vertices has $(N-1)!/2$ permutations, precluding an
exhaustive search except for small $N$. Even a relatively small problem (e.g.,
$N=20$) has $10^{16}$ distinct HCs; $N=40$ leads to $10^{46}$ distinct HCs.
The Euclidean TSP is classified as an NP-hard problem$^{1}$, having no known
algorithm for the general case whose number of operations is a polynomial
function of $N$.

The $(N-1)!/2$ permutations assume that any vertex can occupy any of $N$
positions. Isolating vertices into spatial zones locks each into a limited
range of positions, subject to boundary vertex permutations. \ This falls into
the general area of dynamic programming$^{2,3,4}$.

Partitioning the vertices into sub-problems has been done for the Euclidean
TSP$^{5-10}$. In particular, Arora$^{6}$ obtained a Polynomial Time
Approximation Scheme (PTAS) generating a tour exceeding the optimal length by
no more than a factor of $1+\varepsilon$ in time $N^{O(1/\varepsilon)}$. The
approach involved a bounding square box dissected into squares and shifted
randomly, with restrictions on edge crossings (to specified portals).
\ Mitchell$^{7}$ independently obtained a similar result.

The approach in this paper dissects the problem lengthwise and finds optimal
Hamiltonian Paths (HPs)---paths visiting each vertex once---for the isolated
zones independently of the others. The number of combinations of boundary
vertices determines the number of optimal HPs for each zone. \ Sets of optimal
HPs for each zone (with embedded HPs from previous zones) generate an HC for
the set of vertices. \ When no boundary vertices are omitted, the optimal HC
will contain an optimal HP found from each zone.

This paper illustrates the procedure for a benchmark problem (i.e.,
ATT48$^{11}$) small enough to permit a detailed description of the entire
solution process. The success of the approach depends on limiting the number
of potential boundary vertices and crossing edges. \ In practice, sometimes as
few as two edges will cross a boundary from one zone to another. \ The number
of crossing edges can be increased, if necessary, to improve the solution.

\section{Outline of the Approach}

Following previous approaches$^{5-10}$, the problem is broken down into
subproblems that depend on each other through boundary interactions. The
boundaries have a lengthwise nature, forming (doubly) open-ended zones. Figure
1 shows the separation of the ATT48$^{11}$ vertices into ten zones by means of
nine introduced boundaries, each dissecting the problem lengthwise. Table 1
summarizes the zones and potential boundary vertices for each. Each zone
connects to adjacent zones via a limited number of edges. (An edge is a
straight line connecting two vertices.)

%

\begin{figure}
[ptb]
\begin{center}
\includegraphics[
trim=0.000000in 1.235934in 0.000000in 1.733158in,
height=3.039in,
width=3.838in
]%
{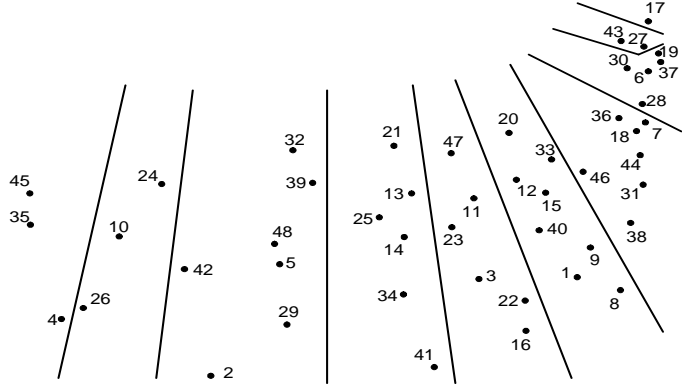}%
\caption{Separation of the ATT48 vertices into ten open-ended zones}%
\end{center}
\end{figure}

A single lengthwise boundary cuts the optimal HC into an even number of HPs,
the sum of which must have the minimum length in each of the two created
spatial zones. For example, if two HPs are created, the HP in each created
zone (terminated at boundary vertices in the other zone) must have the minimum
length. If an HP length exceeds the minimum, replacing it with another HP
(having the same vertices) will reduce the overall HC length. \ Stated another
way, it is not possible to dissect the optimal HC into two HPs and replace one
of them with a shorter HP having the same vertices.

The boundary vertices contained by the optimal HC associated with a particular
dissection are in general not known, requiring the enumeration of all possible
boundary vertices located in the adjacent zone. \ Typically, not all potential
boundary vertices will connect edges to the adjacent zone. For example, as few
as two edges ($n=2$) might connect two zones. \ For each value of $n$, the
binomial coefficient $\left(
\begin{array}
[c]{c}%
b\\
n
\end{array}
\right)  $ provides the number of boundary vertex combinations ($b$ is the
number of potential boundary vertices). \ Summing over all values of $n$ leads
to $2^{b-1}-1$ combinations (when $n=0$ and odd values of $n$ are eliminated).
\ A minimum-length HP is then found for each particular boundary vertex
combination, beginning at the left end zone in figure 1 (zone 1).

The second boundary from the left in figure 1 isolates both zones 1 \& 2 from
the other vertices. The approach then finds the set of minimum length HPs for
the combined vertices in zones 1 \& 2 in the same way, except that the
previously determined HPs from zone 1 become embedded in the new HPs.

Boundary vertices can comprise all the vertices in the adjacent zone, or (more
likely) a smaller subset, usually those closest to the boundary. Vertices
close to the boundary often have the effect of eliminating other potential
boundary vertices because the latter often lead to non-optimal HCs.

\bigskip

Table 1. ATT48 Zones and Boundary Vertices

\bigskip%

\begin{tabular}
[c]{|c|c|c|}\hline
Zone & Vertices & Potential Boundary Vertices\\\hline
$1$ & $4,35,45$ & $26,10,24$\\\hline
$2$ & $26,10,24$ & $2,29,42,48,39,32$\\\hline
$3$ & $2,29,42,5,48,39,32$ & $41,34,14,25,13,21$\\\hline
$4$ & $41,34,14,25,13,21$ & $16,22,3,23,11,47$\\\hline
$5$ & $16,22,3,23,11,47$ & $8,1,9,40,15,12,20$\\\hline
$6$ & $8,1,9,40,15,12,33,20$ & $38,31,46,44,36$\\\hline
$7$ & $38,31,46,44,18,7,36$ & $28,6,30$\\\hline
$8$ & $28,6,30,37,19$ & $27,43$\\\hline
$9$ & $27,43$ & $17$\\\hline
$10$ & $17$ & -----\\\hline
\end{tabular}

\bigskip

The set of minimum-length HPs found for each zone (combined with all
previously-considered zones) includes embedded HPs from the previous zones.
\ However, as the approach determines HPs for later zones, it automatically
begins to filter out non-optimal embedded HPs from previous zones, until at
the last zone, $n=b=2$, and no extraneous HPs remain.

\section{Detailed Description of the Solution}

When the introduced boundaries create zones with boundary vertices confined to
the adjacent zones, the sets of candidate HPs are found by advancing one zone
at a time, considering only the vertices in the zone in question (with
embedded HPs from previous zones) and its adjacent zone to the right.

The zone 1 vertices (4, 35, \& 45) can connect to two of the three boundary
vertices in zone 2 via inter-zone edges according to one of three
combinations: 10 \& 26, 26 \& 24, or 10 \& 24. \ Determination of
minimum-length HPs involves evaluating all interior vertex permutations for
each of the three boundary vertex combinations. \ Table 2 shows the results.

\bigskip

Table 2. Candidate HPs for zone 1

\bigskip%

\begin{tabular}
[c]{|c|c|c|c|c|}\hline
$\ 26$ & $\ \ 4$ & $\ 35$ & $\ 45$ & $\ 10$\\\hline
$\ 26$ & $\ \ 4$ & $\ 35$ & $\ 45$ & $\ 24$\\\hline
$\ 10$ & $\ \ 4$ & $\ 35$ & $\ 45$ & $\ 24$\\\hline
\end{tabular}

\bigskip

Introduction of the second boundary leads to the determination of HPs for the
combined vertices in zone 1 and zone 2 (i.e., vertices 26, 10, \& 24). \ Each
HP terminates to two (or more) of the boundary vertices 2, 29, 42, 48, 39, \&
32. (Vertex 5 is effectively shielded by vertices 48 and 42.) \ When $n=2$,
the six boundary vertices in zone 3 have fifteen possible combinations.
Although $n=4$ is possible, it would require two edges from vertex 10 to cross
the boundary. \ Including extra crossing edges would lead to the evaluation of
more boundary vertex combinations, and would involve determining optimal HPs
on the basis of the sum of their lengths (with embedded HPs from zone 1).

Table 3 shows the possibilities searched in zone 2 for the candidate HPs when
$n=2$. Vertices $V_{1}$ and $V_{2}$ are two of the boundary vertices 2, 29,
42, 48, 39, and 32. \ Embedded HPs (e.g., \textbf{10}-\textbf{24},
\textbf{10}-\textbf{26}, \& \textbf{24}-\textbf{26}) are shown in bold
typeface, both in the text and in the tables.

\bigskip

Table 3. Zone 2 possibilities searched with embedded HPs from zone 1

\bigskip%
\begin{tabular}
[c]{|c|c|c|c|c|c|c|c|c|c|c|c|c|c|}\hline
$V_{1}$ & $V_{1}$ & $V_{1}$ & $V_{1}$ &  & $V_{1}$ & $V_{1}$ & $V_{1}$ &
$V_{1}$ &  & $V_{1}$ & $V_{1}$ & $V_{1}$ & $V_{1}$\\\hline
$\boldsymbol{10}$ & $\boldsymbol{24}$ & $26$ & $26$ &  & $\boldsymbol{26}$ &
$\boldsymbol{24}$ & $10$ & $10$ &  & $\boldsymbol{10}$ & $\boldsymbol{26}$ &
$24$ & $24$\\\hline
$\boldsymbol{24}$ & $\boldsymbol{10}$ & $\boldsymbol{10}$ & $\boldsymbol{24}$
&  & $\boldsymbol{24}$ & $\boldsymbol{26}$ & $\boldsymbol{26}$ &
$\boldsymbol{24}$ &  & $\boldsymbol{26}$ & $\boldsymbol{10}$ &
$\boldsymbol{10}$ & $\boldsymbol{26}$\\\hline
$26$ & $26$ & $\boldsymbol{24}$ & $\boldsymbol{10}$ &  & $10$ & $10$ &
$\boldsymbol{24}$ & $\boldsymbol{26}$ &  & $24$ & $24$ & $\boldsymbol{26}$ &
$\boldsymbol{10}$\\\hline
$V_{2}$ & $V_{2}$ & $V_{2}$ & $V_{2}$ &  & $V_{2}$ & $V_{2}$ & $V_{2}$ &
$V_{2}$ &  & $V_{2}$ & $V_{2}$ & $V_{2}$ & $V_{2}$\\\hline
\end{tabular}

\bigskip

Each of the twelve possibilities in table 3 are searched for the fifteen
$V_{1}$/$V_{2}$ combinations to obtain fifteen minimum-length HPs for zone 2
(table 4), with embedded HPs in bold. \ The zone 2 solution contains only
embedded HPs \textbf{26}-\textbf{10} \& \textbf{26}-\textbf{24}, eliminating
HP \textbf{10}-\textbf{24}.

\bigskip

Table 4. Candidate HPs for zone 2

\bigskip%
\begin{tabular}
[c]{|c|c|c|c|c|}\hline
$2$ & $\boldsymbol{26}$ & $\boldsymbol{10}$ & $24$ & $29$\\\hline
$2$ & $\boldsymbol{26}$ & $\boldsymbol{10}$ & $24$ & $42$\\\hline
$2$ & $\boldsymbol{26}$ & $\boldsymbol{10}$ & $24$ & $48$\\\hline
$2$ & $\boldsymbol{26}$ & $\boldsymbol{10}$ & $24$ & $39$\\\hline
$2$ & $\boldsymbol{26}$ & $\boldsymbol{10}$ & $24$ & $32$\\\hline
$29$ & $\boldsymbol{26}$ & $\boldsymbol{10}$ & $24$ & $42$\\\hline
$29$ & $\boldsymbol{26}$ & $\boldsymbol{10}$ & $24$ & $48$\\\hline
$29$ & $\boldsymbol{26}$ & $\boldsymbol{10}$ & $24$ & $39$\\\hline
$29$ & $\boldsymbol{26}$ & $\boldsymbol{10}$ & $24$ & $32$\\\hline
$42$ & $\boldsymbol{26}$ & $\boldsymbol{10}$ & $24$ & $48$\\\hline
$42$ & $\boldsymbol{26}$ & $\boldsymbol{10}$ & $24$ & $39$\\\hline
$42$ & $\boldsymbol{26}$ & $\boldsymbol{10}$ & $24$ & $32$\\\hline
$48$ & $10$ & $\boldsymbol{26}$ & $\boldsymbol{24}$ & $39$\\\hline
$48$ & $10$ & $\boldsymbol{26}$ & $\boldsymbol{24}$ & $32$\\\hline
$39$ & $10$ & $\mathbf{26}$ & $\mathbf{24}$ & $32$\\\hline
\end{tabular}

\bigskip

The zone 3 solution (table 5) has only two distinct embedded HPs:
\ \textbf{2}-\textbf{42} \& \textbf{32}-\textbf{42}. \ Table 4 shows that both
contain the embedded HP \textbf{26}-\textbf{10} from zone 1.

\bigskip

Table 5. \ Candidate HPs for zone 3

\bigskip%
\begin{tabular}
[c]{|c|c|c|c|c|c|c|c|c|}\hline
$41$ & $29$ & $2$ & $\boldsymbol{42}$ & $\boldsymbol{32}$ & $39$ & $48$ & $5$
& $34$\\\hline
$41$ & $29$ & $\boldsymbol{2}$ & $\boldsymbol{42}$ & $5$ & $48$ & $32$ & $39$
& $14$\\\hline
$41$ & $29$ & $\boldsymbol{2}$ & $\boldsymbol{42}$ & $5$ & $48$ & $32$ & $39$
& $25$\\\hline
$41$ & $29$ & $\boldsymbol{2}$ & $\boldsymbol{42}$ & $5$ & $48$ & $39$ & $32$
& $13$\\\hline
$41$ & $29$ & $\boldsymbol{2}$ & $\boldsymbol{42}$ & $5$ & $48$ & $39$ & $32$
& $21$\\\hline
$34$ & $29$ & $\boldsymbol{2}$ & $\boldsymbol{42}$ & $5$ & $48$ & $32$ & $39$
& $14$\\\hline
$34$ & $29$ & $\boldsymbol{2}$ & $\boldsymbol{42}$ & $5$ & $48$ & $32$ & $39$
& $25$\\\hline
$34$ & $29$ & $\boldsymbol{2}$ & $\boldsymbol{42}$ & $5$ & $48$ & $39$ & $32$
& $13$\\\hline
$34$ & $29$ & $\boldsymbol{2}$ & $\boldsymbol{42}$ & $5$ & $48$ & $39$ & $32$
& $21$\\\hline
$14$ & $48$ & $5$ & $29$ & $2$ & $\boldsymbol{42}$ & $\boldsymbol{32}$ & $39$
& $25$\\\hline
$14$ & $48$ & $5$ & $29$ & $2$ & $\boldsymbol{42}$ & $\boldsymbol{32}$ & $39$
& $13$\\\hline
$14$ & $48$ & $5$ & $29$ & $2$ & $\boldsymbol{42}$ & $\boldsymbol{32}$ & $39$
& $21$\\\hline
$25$ & $48$ & $5$ & $29$ & $2$ & $\boldsymbol{42}$ & $\boldsymbol{32}$ & $39$
& $13$\\\hline
$25$ & $48$ & $5$ & $29$ & $2$ & $\boldsymbol{42}$ & $\boldsymbol{32}$ & $39$
& $21$\\\hline
$13$ & $39$ & $48$ & $5$ & $29$ & $2$ & $\boldsymbol{42}$ & $\boldsymbol{32}$
& $21$\\\hline
\end{tabular}

\bigskip

Zone 4 connects edges to four boundary vertices in zone 5\ (table 6),
generating two HPs for each boundary vertex combination. For each case, either
the 1$^{\text{st}}$ \& 2$^{\text{nd}}$ and 3$^{\text{rd}}$ \& 4$^{\text{th}}$,
or the 1$^{\text{st}}$ \& 4$^{\text{th}}$ and 2$^{\text{nd}}$ \&
3$^{\text{rd}}$ boundary vertices can define the two HPs, effectively doubling
the number of combinations. The number of vertices in each HP can vary, but
must sum to ten, and only the pair that minimizes the sum of their lengths is retained.

Zone 4 contains only six distinct embedded HPs: \textbf{34}-\textbf{21},
\textbf{34}-\textbf{25}, \textbf{25}-\textbf{21}, \textbf{41}-\textbf{21}, and
\textbf{41}-\textbf{25}.

\bigskip

Table 6. \ Candidate HPs for zone 4

\bigskip%
\begin{tabular}
[c]{|c|c|c|c|c|c|c|c|c|c|c|c|c|c|}\hline
\multicolumn{7}{|c|}{1$^{\text{st}}$ HP} & \multicolumn{7}{|c|}{2$^{\text{nd}%
}$ HP}\\\hline
$16$ & $41$ & $22$ &  &  &  &  & $3$ & $\boldsymbol{34}$ & $\boldsymbol{21}$ &
$13$ & $25$ & $14$ & $23$\\\hline
$16$ & $41$ & $22$ &  &  &  &  & $3$ & $\boldsymbol{34}$ & $\boldsymbol{21}$ &
$13$ & $25$ & $14$ & $11$\\\hline
$16$ & $41$ & $22$ &  &  &  &  & $3$ & $\boldsymbol{34}$ & $\boldsymbol{25}$ &
$14$ & $13$ & $21$ & $47$\\\hline
$16$ & $41$ & $34$ & $22$ &  &  &  & $23$ & $14$ & $\boldsymbol{25}$ &
$\boldsymbol{21}$ & $13$ & $11$ & \\\hline
$16$ & $41$ & $22$ &  &  &  &  & $23$ & $13$ & $25$ & $14$ & $\boldsymbol{34}$
& $\boldsymbol{21}$ & $47$\\\hline
$16$ & $41$ & $22$ &  &  &  &  & $11$ & $13$ & $25$ & $14$ & $\boldsymbol{34}$
& $\boldsymbol{21}$ & $47$\\\hline
$16$ & $41$ & $34$ & $3$ &  &  &  & $23$ & $14$ & $\boldsymbol{25}$ &
$\boldsymbol{21}$ & $13$ & $11$ & \\\hline
$16$ & $\boldsymbol{41}$ & $\boldsymbol{21}$ & $47$ &  &  &  & $3$ & $34$ &
$14$ & $25$ & $13$ & $23$ & \\\hline
$16$ & $41$ & $34$ & $3$ &  &  &  & $11$ & $13$ & $14$ & $\boldsymbol{25}$ &
$\boldsymbol{21}$ & $47$ & \\\hline
$16$ & $41$ & $\boldsymbol{34}$ & $\boldsymbol{21}$ & $47$ &  &  & $23$ & $14$
& $25$ & $13$ & $11$ &  & \\\hline
$22$ & $41$ & $34$ & $3$ &  &  &  & $23$ & $14$ & $\boldsymbol{25}$ &
$\boldsymbol{21}$ & $13$ & $11$ & \\\hline
$22$ & $41$ & $34$ & $3$ &  &  &  & $23$ & $13$ & $14$ & $\boldsymbol{25}$ &
$\boldsymbol{21}$ & $47$ & \\\hline
$22$ & $41$ & $34$ & $3$ &  &  &  & $11$ & $13$ & $14$ & $\boldsymbol{25}$ &
$\boldsymbol{21}$ & $47$ & \\\hline
$22$ & $41$ & $\boldsymbol{34}$ & $\boldsymbol{25}$ & $14$ & $23$ &  & $11$ &
$13$ & $21$ & $47$ &  &  & \\\hline
$3$ & $34$ & $\boldsymbol{41}$ & $\boldsymbol{25}$ & $14$ & $23$ &  & $11$ &
$13$ & $21$ & $47$ &  &  & \\\hline
\end{tabular}

\bigskip

Zone 5 (table 7) has only four distinct sets of embedded HPs from zone 4:
\textbf{16}-\textbf{23} \& \textbf{47}-\textbf{11}; \textbf{16}-\textbf{47} \&
\textbf{3}-\textbf{23}; \textbf{16}-\textbf{47} \& \textbf{23}-\textbf{11};
and \textbf{16}-\textbf{23} \& \textbf{11}-\textbf{47}.

Table 6 shows that the four distinct HPs in zone 4 (that are embedded in zone
5) contain only two distinct HPs from zone 3: \textbf{34}-\textbf{21} \&
\textbf{41}-\textbf{21}. Both have the embedded HP \textbf{2}-\textbf{42}.
\ In other words, the approach continues to automatically filter out
extraneous HPs that locally had a minimum length in a previous combination of
zones (for a particular boundary vertex combination), but are not consistent
with the global minimum-length HC.

In table 7, the first HP connects two edges to zone 6. \ The second HP
demonstrates the "closing the loop" process necessary when the number of
crossing edges $n$ decreases from one boundary to the next. \ In this case,
$n$ decreases from four (across the fourth boundary) to two (across the fifth
boundary), and both ends of the 2$^{\text{nd}}$ HP terminate at boundary
vertices in zone 4. \ As already noted, the terminating loop can contain ends
from two separate HPs from zone 4. \ For example, the HPs \textbf{16}%
-\textbf{47} \& \textbf{11}-\textbf{23} from zone 4 lead to HPs \textbf{16}%
-\textbf{23} and \textbf{47}-\textbf{11} in zone 5; HP \textbf{16}-\textbf{23}
terminates in zone 5 when $n=2$ in zone 6.

Zones 6 to 8 have only two edges connecting to either adjacent zone. \ The
only remaining embedded HPs in zone 6 (table 8) are \textbf{12}-\textbf{20} \&
\textbf{1}-\textbf{20}, reducing the embedded HPs from zone 5 to
\textbf{11}-\textbf{47} \& \textbf{16}-\textbf{23}, and \textbf{16}%
-\textbf{47} \& \textbf{23}-\textbf{11}.

\bigskip

Table 7. \ Candidate HPs for zone 5

\bigskip%
\begin{tabular}
[c]{|c|c|c|c|c|c|c|c|c|c|c|}\hline
\multicolumn{6}{|c|}{1$^{\text{st}}$ HP} & \multicolumn{5}{|c|}{2$^{\text{nd}%
}$ HP}\\\hline
$8$ & $\boldsymbol{16}$ & $\boldsymbol{23}$ & $3$ & $22$ & $1$ &  &
$\boldsymbol{47}$ & $\boldsymbol{11}$ &  & \\\hline
$8$ & $22$ & $\boldsymbol{16}$ & $\boldsymbol{23}$ & $3$ & $9$ &  &
$\boldsymbol{47}$ & $\boldsymbol{11}$ &  & \\\hline
$8$ & $22$ & $\boldsymbol{16}$ & $\boldsymbol{23}$ & $3$ & $40$ &  &
$\boldsymbol{47}$ & $\boldsymbol{11}$ &  & \\\hline
$8$ & $22$ & $\boldsymbol{16}$ & $\boldsymbol{47}$ & $11$ & $15$ &  &
$\boldsymbol{3}$ & $\boldsymbol{23}$ &  & \\\hline
$8$ & $22$ & $\boldsymbol{16}$ & $\boldsymbol{47}$ & $11$ & $12$ &  &
$\boldsymbol{3}$ & $\boldsymbol{23}$ &  & \\\hline
$8$ & $22$ & $3$ & $\boldsymbol{16}$ & $\boldsymbol{47}$ & $20$ &  &
$\boldsymbol{23}$ & $\boldsymbol{11}$ &  & \\\hline
$1$ & $22$ & $\boldsymbol{16}$ & $\boldsymbol{23}$ & $3$ & $9$ &  &
$\boldsymbol{47}$ & $\boldsymbol{11}$ &  & \\\hline
$1$ & $22$ & $\boldsymbol{16}$ & $\boldsymbol{23}$ & $3$ & $40$ &  &
$\boldsymbol{47}$ & $\boldsymbol{11}$ &  & \\\hline
$1$ & $22$ & $\boldsymbol{16}$ & $\boldsymbol{47}$ & $11$ & $15$ &  &
$\boldsymbol{3}$ & $\boldsymbol{23}$ &  & \\\hline
$1$ & $22$ & $\boldsymbol{16}$ & $\boldsymbol{47}$ & $11$ & $12$ &  &
$\boldsymbol{3}$ & $\boldsymbol{23}$ &  & \\\hline
$1$ & $3$ & $22$ & $\boldsymbol{16}$ & $\boldsymbol{47}$ & $20$ &  &
$\boldsymbol{23}$ & $\boldsymbol{11}$ &  & \\\hline
$9$ & $22$ & $\boldsymbol{16}$ & $\boldsymbol{23}$ & $3$ & $40$ &  &
$\boldsymbol{47}$ & $\boldsymbol{11}$ &  & \\\hline
$9$ & $3$ & $22$ & $\boldsymbol{16}$ & $\boldsymbol{47}$ & $15$ &  &
$\boldsymbol{23}$ & $\boldsymbol{11}$ &  & \\\hline
$9$ & $3$ & $22$ & $\boldsymbol{16}$ & $\boldsymbol{47}$ & $12$ &  &
$\boldsymbol{23}$ & $\boldsymbol{11}$ &  & \\\hline
$9$ & $3$ & $22$ & $\boldsymbol{16}$ & $\boldsymbol{47}$ & $20$ &  &
$\boldsymbol{23}$ & $\boldsymbol{11}$ &  & \\\hline
$40$ & $3$ & $22$ & $\boldsymbol{16}$ & $\boldsymbol{47}$ & $15$ &  &
$\boldsymbol{23}$ & $\boldsymbol{11}$ &  & \\\hline
$40$ & $3$ & $22$ & $\boldsymbol{16}$ & $\boldsymbol{47}$ & $12$ &  &
$\boldsymbol{23}$ & $\boldsymbol{11}$ &  & \\\hline
$40$ & $3$ & $22$ & $\boldsymbol{16}$ & $\boldsymbol{47}$ & $20$ &  &
$\boldsymbol{23}$ & $\boldsymbol{11}$ &  & \\\hline
$15$ & $\boldsymbol{11}$ & $\boldsymbol{47}$ & $12$ &  &  &  &
$\boldsymbol{16}$ & $22$ & $3$ & $\boldsymbol{23}$\\\hline
$15$ & $\boldsymbol{11}$ & $\boldsymbol{47}$ & $20$ &  &  &  &
$\boldsymbol{16}$ & $22$ & $3$ & $\boldsymbol{23}$\\\hline
$12$ & $\boldsymbol{11}$ & $\boldsymbol{47}$ & $20$ &  &  &  &
$\boldsymbol{16}$ & $22$ & $3$ & $\boldsymbol{23}$\\\hline
\end{tabular}

\bigskip

\bigskip

Table 8. Candidate HPs for zone 6

\bigskip%
\begin{tabular}
[c]{|c|c|c|c|c|c|c|c|c|c|}\hline
$38$ & $8$ & $1$ & $9$ & $40$ & $15$ & $\boldsymbol{12}$ & $\boldsymbol{20}$ &
$33$ & $31$\\\hline
$38$ & $8$ & $1$ & $9$ & $40$ & $15$ & $\boldsymbol{12}$ & $\boldsymbol{20}$ &
$33$ & $46$\\\hline
$38$ & $8$ & $1$ & $9$ & $40$ & $15$ & $\boldsymbol{12}$ & $\boldsymbol{20}$ &
$33$ & $36$\\\hline
$31$ & $9$ & $8$ & $1$ & $40$ & $15$ & $\boldsymbol{12}$ & $\boldsymbol{20}$ &
$33$ & $46$\\\hline
$31$ & $9$ & $8$ & $1$ & $40$ & $15$ & $\boldsymbol{12}$ & $\boldsymbol{20}$ &
$33$ & $36$\\\hline
$46$ & $33$ & $12$ & $15$ & $40$ & $9$ & $8$ & $\boldsymbol{1}$ &
$\boldsymbol{20}$ & $36$\\\hline
\end{tabular}

\bigskip

Table 9 indicates that zone 7 has only one embedded HP (\textbf{38-46}), which
has the effect of eliminating all extraneous (non-optimal) embedded HPs from
all previous zones.

Finally, both zones 8 and 9 have only two edges crossing their right
boundaries ($n=b=2$), reducing the number of minimum-length HPs in both cases
to one: \ 27-19-37-6-\textbf{28}-\textbf{30}-43 (zone 8), and 17-\textbf{27}%
-\textbf{43}-17 (zone 9). \ 

The optimal HC (table 10) results from working backwards to extract the
embedded HP (\textbf{28-30}) from the zone 8 solution, and then extracting the
embedded HP (\textbf{38-46}), from the zone 7 solution, etc. Substituting all
of the vertices into embedded HPs leads to the overall solution.

\bigskip

\bigskip

Table 9. Candidate HPs for zone 7

\bigskip%
\begin{tabular}
[c]{|c|c|c|c|c|c|c|c|c|}\hline
$28$ & $7$ & $18$ & $44$ & $31$ & $\boldsymbol{38}$ & $\boldsymbol{46}$ & $36$
& $6$\\\hline
$28$ & $7$ & $18$ & $44$ & $31$ & $\boldsymbol{38}$ & $\boldsymbol{46}$ & $36$
& $30$\\\hline
$28$ & $7$ & $18$ & $44$ & $31$ & $\boldsymbol{38}$ & $\boldsymbol{46}$ & $36$
& $37$\\\hline
$28$ & $7$ & $18$ & $44$ & $31$ & $\boldsymbol{38}$ & $\boldsymbol{46}$ & $36$
& $19$\\\hline
$6$ & $7$ & $18$ & $44$ & $31$ & $\boldsymbol{38}$ & $\boldsymbol{46}$ & $36$
& $30$\\\hline
$6$ & $36$ & $\boldsymbol{46}$ & $\boldsymbol{38}$ & $31$ & $44$ & $18$ & $7$
& $37$\\\hline
$6$ & $36$ & $\boldsymbol{46}$ & $\boldsymbol{38}$ & $31$ & $44$ & $18$ & $7$
& $19$\\\hline
$30$ & $36$ & $\boldsymbol{46}$ & $\boldsymbol{38}$ & $31$ & $44$ & $18$ & $7$
& $37$\\\hline
$30$ & $36$ & $\boldsymbol{46}$ & $\boldsymbol{38}$ & $31$ & $44$ & $18$ & $7$
& $19$\\\hline
$37$ & $7$ & $18$ & $44$ & $31$ & $\boldsymbol{38}$ & $\boldsymbol{46}$ & $36$
& $19$\\\hline
\end{tabular}

\bigskip

Table 10. \ ATT48 Solution

\bigskip%

\begin{tabular}
[c]{|c|c|c|c|c|c|c|c|c|c|c|c|}\hline
Zone & \multicolumn{11}{|c|}{Optimal HPs (embedded HPs in bold)}\\\hline
$1$ & $26$ & $4$ & $35$ & $45$ & $10$ &  &  &  &  &  & \\\hline
$2$ & $2$ & $\boldsymbol{26}$ & $\boldsymbol{10}$ & $24$ & $42$ &  &  &  &  &
& \\\hline
$3$ & $34$ & $29$ & $\boldsymbol{2}$ & $\boldsymbol{42}$ & $5$ & $48$ & $39$ &
$32$ & $21$ &  & \\\hline
$4$ & $16$ & $41$ & $\boldsymbol{34}$ & $\boldsymbol{21}$ & $47$ &  & $23$ &
$14$ & $25$ & $13$ & $11$\\\hline
$5$ & $12$ & $\boldsymbol{11}$ & $\boldsymbol{47}$ & $20$ &  & $16$ & $22$ &
$3$ & $23$ &  & \\\hline
$6$ & $38$ & $8$ & $1$ & $9$ & $40$ & $15$ & $\boldsymbol{12}$ &
$\boldsymbol{20}$ & $33$ & $46$ & \\\hline
$7$ & $28$ & $7$ & $18$ & $44$ & $31$ & $\boldsymbol{38}$ & $\boldsymbol{46}$
& $36$ & $30$ &  & \\\hline
$8$ & $27$ & $19$ & $37$ & $6$ & $\boldsymbol{28}$ & $\boldsymbol{30}$ & $43$
&  &  &  & \\\hline
$9$ & $17$ & $\boldsymbol{27}$ & $\boldsymbol{43}$ & $17$ &  &  &  &  &  &  &
\\\hline
\end{tabular}

\bigskip

\section{Concluding Remarks}

Introducing lengthwise boundaries allows optimal HPs to be determined locally
for each zone (one for each boundary vertex combination), and also allows the
solution to progress successively from zone to zone, automatically filtering
out previous HPs that are inconsistent with a globally minimum-length HC.
\ Embedded HPs from previous zones helps to reduce the computation time.

The solution efficiency depends on the number of boundary vertices and
crossing edges for each zone. \ ATT48 requires only two inter-zone edges from
each zone, except zone 4, which has four inter-zone edges (to zone 5).
\ Although the approach considered only limited values of $n$ and $b$ rather
than all possible values, the approach can also increase $n$ and $b$ for more
complex problems.

Including results for $n\neq2$ (or $n\neq4$ for zone 4) will add non-optimal
solutions to ATT48, increasing the computation time linearly with the added
number of combinations. \ The limited boundary vertex combinations considered
required 430 seconds of CPU time. Considering all boundary vertex combinations
increases the estimated run time by a factor of approximately 36.

\section{References}

\begin{enumerate}
\item C. H. Papadimitriou (1977). \ Euclidean TSP is NP-complete.
\ \textit{Theoretical Computer Science} \textbf{4}, 237-244.

\item R. S. Bellman \& S. Dreyfus (1962). \ \textit{Applied Dynamic
Programming.} \ Princeton Univ. Press, Princeton, N.J.

\item R. E. Larson \& J. L. Casti (1982). \ \textit{Principles of Dynamic
Programming, Vol. I, II}. \ Marcel Dekker, New York.

\item T. H. Cormen, C. E. Leiserson, R. L. Rivest, \& C. Stein (2001).
\ \textit{Introduction to Algorithms}. \ MIT Press \& McGraw-Hill, p. 323-369.

\item R. M. Karp (1977). Probabilistic analysis of partitioning algorithms for
the TSP in the plane. \textit{Math. Oper. Res.} \textbf{2}, 209-224.

\item S. Arora (1996). Polynomial time approximation schemes for Euclidean TSP
and other geometric problems, \textit{Proc. 37th Annual Sympos. on Foundations
of Computer Science}, p.2.

\item J. S. B. Mitchell (1996). Guillotine subdivisions approximate polygonal
subdivisions: a simple new method for the geometric k-MST problem.
\textit{Proc. 7th ACM-SIAM Sympos. on Discrete Algorithms}, p. 402.

\item E. L. Lawler \textit{et al.} (1985). \ \textit{The Traveling Salesman
Problem}. \ Wiley, p. 185.

\item S. Arora (2003). \ Approximation schemes for NP-hard geometric
optimization problems: \ a survey. \ \textit{Math. Program., Ser. B}
\textbf{97}, 43-69.

\item G. Cesari (1996). \ Divide and conquer strategies for parallel TSP
heuristics. \ \textit{Computers Ops. Res.} \textbf{7}, 681-684.

\item http://www.iwr.uni-heidelberg.de/groups/comopt/software/TSPLIB95/tsp/.
\end{enumerate}

\end{document}